\documentclass{article}
\usepackage{amsmath}
\usepackage{amssymb}
\usepackage{graphicx}
\begin{document}
\begin{center}
\Large
\textbf{Bit symmetry entails the symmetry of the quantum transition probability}
\vspace{0,3 cm}
\normalsize

Gerd Niestegge 
\footnotesize
\vspace{0,2 cm}

Ahaus, Germany

gerd.niestegge@web.de, https://orcid.org/0000-0002-3405-9356
\end{center}
\normalsize
\begin{abstract}
It is quite common to use the 
generalized probabilistic theories (GPTs)
as generic models 
to reconstruct quantum theory from a few basic principles
and to gain
a better understanding of the 
probabilistic or information theoretic 
foundations of quantum physics
and quantum computing.
A variety of symmetry postulates 
was introduced and studied
in this framework, 
including the transitivity of the automorphism group 
(1) on the pure states,
(2) on the pairs of orthogonal pure states 
[these pairs are called 2-frames] and
(3) on any frames of the same size.
The second postulate is M\"uller and Ududec's \emph{bit symmetry},
which they motivate by quantum computational needs.
Here we explore these three postulates in the 
transition probability framework, 
which is more specific than the GPTs
since the existence of the transition probabilities
for the quantum logical atoms is presupposed 
either directly or indirectly 
via a certain geometric property of the state space. 
This property for compact convex sets was introduced by the author in a recent paper.
We show that bit symmetry implicates the symmetry of the transition probabilities
between the atoms.
Using a result by Barnum and Hilgert, we 
can then conclude that the third rather strong symmetry postulate 
rules out all models but the classical cases and
the simple Euclidean Jordan algebras.
\vspace{0,3 cm}

\noindent
\textbf{Keywords:} 
quantum transition probability; 
convex sets; 
state spaces;
quantum information;
self-dual cones;
Euclidean Jordan algebras; 
quantum logics
\end{abstract}

\section{Introduction}

Generalized probabilistic theories (GPTs \cite{Barnum2020composites, 
barrett2007information, muller2021probabilistic, muller2012ududec, 
Wilce2019conjugatesfilters})
are commonly used as generic models 
to reconstruct quantum theory from a few basic principles
and to gain
a better understanding of the 
probabilistic and information theoretic 
foundations of quantum physics
and quantum computing.
A more specific model is
the author's transition probability 
framework 
\cite{nie2020alg_origin, nie2021generic, Nie2022genqubit, nie2023conv_self-dual}. 
A distinguishing feature is
the direct or indirect postulate
that the transition probability
must exist for the quantum logical atoms
or, equivalently, that there is one unique state 
for each atom in which the atom carries 
the probability 1. 
This equivalent postulate is sometimes called
\emph{sharpness} \cite{Wilce2019conjugatesfilters}.

A single geometric property of a compact convex set
that gives rise to such a model
was discovered in Ref. \cite{nie2023conv_self-dual}.
It is presupposed here as a given fact and 
we refer to Ref. \cite{nie2023conv_self-dual}
for background information and motivating considerations.
\newpage

We use this model and the geometric property to study
several well-known symmetry postulates of the GPTs
and particularly M\"uller and Ududec's \emph{bit symmetry}, 
which they motivate by quantum computational needs
\cite{muller2012ududec}.
We show that bit symmetry implies the
symmetry of the transitions probabilities between the atoms. 
Using a result by Barnum and Hilgert~\cite{BarnumHilgert2020}, we 
can then conclude that a stronger form of symmetry 
rules out all models but the classical cases and
the simple Euclidean Jordan algebras.

The transition probability framework and some results from
previous papers (particularly \cite{nie2023conv_self-dual}) 
that will be needed here are briefly recapitulated 
in the next section. The symmetry postulates under consideration
are defined and discussed in section 3. Section 4 is dedicated to the 
first and most weak one. In section~5 bit symmetry is studied and 
our main result (Theorem 1) is presented.
The strong form of symmetry is considered in section~6. 

\section{A brief synopsis of the transition probability framework}

Let $\Omega$ be any compact convex set in some 
finite-dimensional real vector space and let
$A_\Omega$ denote the order unit space that consists of the
\emph{affine} real-valued functions on $\Omega$; its order unit
is the constant function $\mathbb{I} \equiv 1$.
The state space of $A_\Omega$ consists of the
positive linear functionals $\mu$ on $A_\Omega$ with $\mu(\mathbb{I})=1$
and is isomorphic to the convex set $\Omega$ 
with the mapping $\omega \rightarrow \delta_\omega$, $\omega \in \Omega$,
where $\delta_\omega (a) := a(\omega)$ for $a \in A_\Omega$.
We consider the following compact convex set in $A_\Omega$
$$\left[0,\mathbb{I}\right] := \left\{a \in A_\Omega | 0 \leq a \leq \mathbb{I}\right\} 
= \left\{a \in A_\Omega | 0 \leq a, \left\|a\right\| \leq 1 \right\}$$
and the set of its extreme points $ext(\left[0,\mathbb{I}\right])$.
For each $\omega \in \Omega$ we define the following function $e_\omega$ on $\Omega$:
$$e_\omega (\zeta) := inf\left\{a(\zeta) : a \in A_\Omega, 
0 \leq a \text{ and } a(\omega)=1 \right\}$$
for $\zeta \in \Omega$. 
Since $\mathbb{I}(\omega) = 1$,
we have $e_\omega (\zeta) \leq 1$ for all $\zeta \in \Omega$.
Generally, $e_\omega$ is not affine
and does not belong to $A_\Omega$. The following novel property of a 
compact convex set $\Omega$ was introduced in Ref. \cite{nie2023conv_self-dual}:
\itshape
\begin{enumerate}
	\item [($\ast \ast$)] For each extreme point $\omega \in \Omega$, 
the function $e_\omega$ is affine (this means $e_\omega \in A_\Omega$) 
and $e_\omega(\zeta) \neq 1$ for all $\zeta \in \Omega$ with $\zeta \neq\omega$.
\end{enumerate}
\normalfont
This means that there is a smallest non-negative affine function 
with the value $1$ at the extreme point and at no other point.
If the condition ($\ast \ast$) holds, then $A_\Omega$ 
has the following properties \cite{nie2023conv_self-dual}:
\begin{enumerate}
\item[(a)]
The set of extreme points
$ext(\left[0,\mathbb{I}\right])$ is an atomic orthomodular lattice
with the orthocomplementation $p \rightarrow p' := \mathbb{I} - p$
for $p \in ext(\left[0,\mathbb{I}\right])$. This set thus becomes a \emph{quantum logic};
its \emph{atoms} are the minimal non-zero elements.
\item[(b)]
For each atom $e$ in $ext(\left[0,\mathbb{I}\right])$
there is one unique state $\mathbb{P}_e$ with $\mathbb{P}_e(e) = 1$.
This state is pure (an extreme point of the state space).
\item[(c)]
For each pure state $\mu$ there is an atom $e$ with $\mu = \mathbb{P}_e$.
\item[(d)]
Two atoms $e_1$ and $e_2$ are \emph{orthogonal} 
if one of the following three equivalent conditions holds:
(1) $e_1 + e_2 \leq \mathbb{I}$,
(2) $\mathbb{P}_{e_1}(e_2) = 0$,
(3) $\mathbb{P}_{e_2}(e_1) = 0$.
Note that the orthogonality of atoms implies their linear independence.
\item[(e)]
$A_\Omega$ is \emph{spectral}; this means that
each $a \in A_\Omega$ can be represented as 
$ a = \sum^{n}_{k=1} s_k e_k $ with
$s_k \in \mathbb{R}$ and pairwise orthogonal
atoms $e_k \in ext(\left[0,\mathbb{I}\right])$,
$k = 1,2,...,n$, $n \in \mathbb{N}$. 
Here we have $0 \leq a$ iff $0 \leq s_k$ for each $k = 1,..., n$. 
\end{enumerate}

Property (b), which is sometimes 
called \emph{sharpness} \cite{Wilce2019conjugatesfilters},
means that the \emph{transition probability} \cite{nie2021generic}
exists for each atom. It is invariant under the automorphisms
$U$ of the order-unit space $A_\Omega$: 
$\mathbb{P}_{e}(a) = \mathbb{P}_{Ue}(Ua)$
for any atom $e$ and any element $a$ in $A_\Omega$ \cite{nie2021generic}.
The transition probability is called \emph{symmetric} if
$\mathbb{P}_{e_2}(e_1) = \mathbb{P}_{e_1}(e_2)$
holds for each pair of atoms $e_1$ and $e_2$.

Following the notation of Ref.~\cite{Barnum2020composites},
the maximum number $m$ of pairwise orthogonal atoms 
is called the \emph{information capacity}; $m$ is finite, 
since $A_\Omega$ has a finite dimension. Then
$\mathbb{I} = e_1 + ... + e_m$ with some pairwise orthogonal
atoms $e_1, ..., e_m$ and the sum of any $m$ 
pairwise orthogonal atoms equals $\mathbb{I}$.
Furthermore 
we have $n \leq m$ for the number $n$ 
in the spectral decomposition (e).

First examples of compact convex sets with ($\ast \ast$) are
the strictly convex and smooth compact convex sets 
\cite{Nie2022genqubit}; they all have 
information capacity $m = 2$. Here the transition probability is 
symmetric only for the Euclidean unit balls.

Further examples are the state spaces of the 
finite-dimensional Euclidean (formally real)
Jordan algebras \cite{AS02, hanche1984jordan, nie2023conv_self-dual}. 
Their transition probabilities are always symmetric 
and they include 
the finite-dimensional classical state spaces (simplexes)
as well as the state 
spaces of finite-dimensional quantum theory.

A more exotic example is the \emph{triangular pillow} \cite{AS02}.
Its information capacity is $m=3$, its transition probabilities
are not symmetric \cite{Nie2022genqubit} and it 
satisfies~($\ast \ast$)~\cite{nie2023conv_self-dual}.

The mathematical property ($\ast \ast$) 
appears to emerge without any motivation here. However, it
does follow from four physically more plausible conditions~\cite{nie2023conv_self-dual}; 
these are the above conditions (b), (c) and (e) 
together with a further condition that is not discussed here and 
that more closely ties the order relation 
to the state space~\cite{nie2023conv_self-dual}.

\section{The symmetry postulates}

We consider the automorphism groups $Aut(\Omega)$ and $Aut(A_\Omega)$
of $\Omega$ and $A_\Omega$, repectively; $Aut(\Omega)$
consists of the bijective affine transformations of $\Omega$ 
and $Aut(A_\Omega)$ consists of the bijective positive linear transformations $U$
of $A_\Omega$ with a positive inverse and $U(\mathbb{I})=\mathbb{I}$.
Each one is a compact group \cite{horne1978autgroupofcone, kimura2014affine}.

For each $T \in Aut(\Omega)$ we get $T^{*} \in Aut(A_\Omega)$ by defining
$(T^{*}a) (\omega):= a(T \omega)$ for $a \in A_\Omega$ and $\omega \in \Omega$.
Vice versa, for each $U \in Aut(A_\Omega)$ we get $U^{*} \in Aut(\Omega)$
by defining $U^{*}\omega := \theta$, 
where $\theta$ is the element in $\Omega$  
with $\delta_\theta = \delta_\omega U$.
\newpage

The first symmetry postulate becomes that $Aut(\Omega)$
acts transitively on the extreme points of $\Omega$
\cite{kimura2014affine, Sanyal2011orbitobes}:
For any two extreme points $\omega_1$ and $\omega_2$
there shall be $T \in Aut(\Omega)$ with $T\omega_1 = T\omega_2$.
Because of (b) and (c) in section 2 this is equivalent
to the property that
$Aut(A_\Omega)$ acts transitively on the atoms 
in $ext(\left[0,\mathbb{I}\right])$. This property
is often called \emph{transitivity}, but here it shall be named
\emph{weak symmetry}, which better matches our terminology of
symmetry conditions.

We say that the \emph{exchange symmetry} holds if
there is $T \in Aut(\Omega)$ with $T\omega_1 = T\omega_2$ and $T\omega_2 = T\omega_1$
for any two extreme points $\omega_1$ and $\omega_2$ in $\Omega$.
This is equivalent
to the property that there is 
$U \in Aut(A_\Omega)$ with $Up = q$ and $Uq=p$ for any two atoms $p$ and $q$
in $ext(\left[0,\mathbb{I}\right])$. 

We call a pair of extreme points $\omega_1$ and $\omega_2$ in $\Omega$
\emph{orthogonal} if the atoms belonging to the corresponding 
pure states are orthogonal. In the generalized probabilistic
theories, the term 
"\emph{perfectly distinguishable}" 
\cite{Barnum2020composites, BarnumHilgert2020, muller2012ududec} 
is often used in this case. 
The set $\Omega$ is called \emph{bit-symmetric} \cite{muller2012ududec}
if there is a transformation $T \in Aut(\Omega)$ with 
$T \omega_1 = T \theta _1$ and $T \omega_2 = T \theta _2$,
whenever $\omega_1$ and $\omega_2$ form any orthogonal pair of extreme points
and $\theta_1$ and $\theta_2$ form any further 
orthogonal pair of extreme points in $\Omega$.
This becomes equivalent to the condition
that any two orthogonal pairs of atoms in $ext(\left[0,\mathbb{I}\right])$
can be mapped to each other by a transformation in $Aut(A_\Omega)$.
Bit symmetry is considered a quantum computational requirement,
since it is thought that a quantum computer must be capable to reversibly transfer
any logical bit to any other logical bit.

\emph{Strong symmetry} \cite{BarnumHilgert2020} means 
that any family of pairwise orthogonal extreme points of $\Omega$
(equivalently, atoms in $ext(\left[0,\mathbb{I}\right])$)
can be mapped to any other such family with the same number of elements
by a transformation in $Aut(\Omega)$ (or $Aut(A_\Omega)$).
Such families are called \emph{frames} in the GPTs 
\cite{Barnum2020composites, BarnumHilgert2020, Wilce2019conjugatesfilters}.

Obviously, the strong symmetry implies the bit symmetry
which again implies weak symmetry.
Moreover, the exchange symmetry implies the weak symmetry,
but its relation to the bit symmetry
and the strong symmetry is not clear.

An immediate important consequence
of the exchange symmetry is 
that the transition probability becomes symmetric.
For any two atoms $e_1$ and $e_2$ we have an 
automorphism $U$ with $Ue_1 = e_2$ and $Ue_2 = e_1$
and then 
$\mathbb{P}_{e_1}(e_2) = \mathbb{P}_{Ue_1}(Ue_2) = \mathbb{P}_{e_2}(e_1)$.
As shown in Ref. \cite{nie2023conv_self-dual} the symmetric transition probability
then results in an inner product $\left\langle \ |\ \right\rangle$ 
on $A_\Omega$ with a self-dual cone and
$\mathbb{P}_{e_1}(e_2) = \left\langle e_1|e_2\right\rangle$
for any atoms $e_1$ and $e_2$.
The main result of this paper will be that 
bit symmetry has the same consequences, but this 
is somewhat more difficult to show.

The finite-dimensional Euclidean Jordan algebras do not
generally satisfy the above symmetry postulates. 
Bit symmetry and
strong symmetry hold only in the 
simple (or non-decomposable or irreducible) algebras
and in the abelian algebras. Weak symmetry 
and exchange symmetry
hold also in the decomposable algebras, if they are direct sums of 
\emph{isomorphic} factors, but do not hold in direct sums of 
non-isomorphic factors.
\newpage

Note that, with information capacity $m=2$, 
weak symmetry, bit symmetry and strong symmetry
become immediately equivalent. If $U p = U q$
holds for two atoms $p$ and $q$ and $U \in Aut(A_\Omega)$,
then $U(\mathbb{I} - p) = U(\mathbb{I} - q)$, but any
orthogonal pair of atoms consists of an atom $e$ 
and the atom $e' = \mathbb{I} - e$ in this case.
Moreover frames with more than two elements do not exist.

Several further symmetry postulates 
for the compact convex sets and the GPTs
can be found in the 
mathematical and physical literature, but
do not play any role here.

\section{Weak symmetry}

\noindent
\textbf{Lemma 1:} 
\itshape
Let $\Omega$ be any finite-dimensional weakly symmetric compact convex set 
with the property ($\ast \ast$) and information capacity $m$.
\begin{enumerate}
\item[(i)]
There exists an $Aut(A_\Omega)$-invariant 
state $\mu_{inv}$ on $A_\Omega$ 
and $\mu_{inv}(e) = 1/m$ for every atom $e$
in $ext(\left[0,\mathbb{I}\right])$.
\item[(ii)]
There  exists an $Aut(A_\Omega)$-invariant inner product 
$\left\langle \  | \ \right\rangle_o$ on $A_\Omega$ with \linebreak
$\left\langle e | e \right\rangle_o = 1$ for every atom $e$.
\item[(iii)]
If $\mathbb{I} = q_1 + ... + q_n$ holds 
with pairwise orthogonal atoms $q_1, ..., q_n$,
then $n = m$.
\end{enumerate}
\vspace{0,3 cm}
\normalfont

Proof. 
(i) and (iii):
Using the normalized Haar measure on $Aut(A_\Omega)$
and arbitrarily selecting any state $\mu_o$,
we define 
\begin{center}
$ \mu_{inv}(a) := \int_{U \in Aut(A_\Omega)} \mu_o(Ua) dU$
\end{center}
for $a \in A_\Omega$. This becomes an $Aut(A_\Omega)$-invariant state.
Since $Aut(A_\Omega)$ acts transitively on the atoms,
we get $\mu_{inv}(p) = \mu_{inv}(q)$ for any atoms $p$ and $q$.

There are $m$ pairwise orthogonal 
atoms $e_1, ..., e_m$ with $\mathbb{I} = e_1 + ... + e_m$.
Then $1 = \mu_{inv}(\mathbb{I}) = \mu_{inv}(e_1) + ... + \mu_{inv}(e_m) = m \mu(p)$ 
and thus $\mu(p)=1/m$ for every atom $p$.
So far we have (i).
Now suppose $\mathbb{I} = q_1 + ... + q_n$
with any further pairwise orthogonal atoms $q_1, ..., q_n$.
Then $1 = \mu_{inv}(\mathbb{I}) = \mu_{inv}(q_1) + ... + \mu_{inv}(q_n) = n/m$.
Therefore $n = m$ and we have (iii).

(ii) 
Let $( \ | \ )$ be any inner product on $A_\Omega$. 
Using again the normalized Haar measure on $Aut(A_\Omega)$,
we construct a further $Aut(A_\Omega)$-invariant inner product via
\begin{center}
$ \left\langle a|b\right\rangle_o := \int_{U \in Aut(A_\Omega)} ( Ua|Ub ) dU $
\end{center}
for any $a,b \in A_\Omega$.
Since $Aut(A_\Omega)$ acts transitively on the atoms,
we have 
$\left\langle p|p \right\rangle_o = \left\langle q|q \right\rangle_o$ 
for any two atoms $p$ and $q$ and we can normalize 
$\left\langle \ | \  \right\rangle_o$ in such a way that 
$\left\langle e|e \right\rangle_o = 1$ for every atom $e$.
\hfill $\square$
\vspace{0,3 cm}

Lemma 1 and its proof constitute a small piece of a proof in 
M\"uller and Ududec's paper \cite{muller2012ududec}. The same thing 
has been shown by Wilce in another way~\cite{Wilce2012four}.
\newpage
\section{Bit symmetry}

So far there is no connection between the orthogonality in the quantum logic
$ext(\left[0,\mathbb{I} \right])$
and the orthogonality in the Euclidean space 
$A_\Omega$ with the inner product $\left\langle \  | \ \right\rangle_o$
from Lemma 1 (ii).
Here we will construct a further inner product with such a connection, 
preassuming bit symmetry and using the state $\mu_{inv}$ and the 
inner product $\left\langle \ | \ \right\rangle_o$ from Lemma 1.
\vspace{0,3 cm}

\noindent
\textbf{Theorem 1:} 
\itshape
Let $\Omega$ be any bit-symmetric finite-dimensional compact convex set
with the property ($\ast \ast$) and information capacity $m$. 

Then $A_\Omega$ possesses an inner product $\left\langle \  | \ \right\rangle$
such that the positive cone becomes self-dual
and $\mathbb{P}_p(q) = \left\langle p|q \right\rangle$ 
holds for any atoms $p$ and $q$.
Therefore the transition probability is symmetric: $\mathbb{P}_p(q) = \mathbb{P}_q(p)$. 

Furthermore,
$ \left\langle \mathbb{I}|x \right\rangle 
= \left\langle x | \mathbb{I} \right\rangle = m \mu_{inv}(x) $ 
for $x \in A_\Omega$.
The atoms are the extreme points of 
the set $\left\{ a \in A_\Omega | 0 \leq a, \mu_{inv}(a) = 1/m \right\}$
and this set is affinely isomorphic 
to the state space of $A_\Omega$ as well as to $\Omega$ itself.
\normalfont
\vspace{0,3 cm}

Proof.
We use the state $\mu_{inv}$ and 
the $Aut(A_\Omega)$-invariant inner product $\left\langle \ | \ \right\rangle_o$ 
from Lemma 1. The information capacity is again denoted by $m$. 
From the bit symmetry we get a real number $\epsilon$ such that
$\epsilon = \left\langle p | q \right\rangle_o $
for any atoms $p, q$ 
with $p + q \leq \mathbb{I}$.
Together with Lemma 1 (ii) the Cauchy-Schwarz inequality implicates 
$\left|\epsilon\right| \leq 1$. Moreover,
the case $ \left|\epsilon\right| = 1$ is impossible,
since atoms $p,q$ with $p + q \leq \mathbb{I}$ are linearly independent,
and we thus have $\left|\epsilon\right| < 1$.
We now define 
$$ \left\langle a|b \right\rangle := \frac{1}{1-\epsilon}\left[\left\langle a|b\right\rangle_o - m^{2} \epsilon \mu_{inv}(a) \mu_{inv}(b)\right]$$
for $a,b \in A_\Omega$.
Then $\left\langle p|p\right\rangle = 1$ for each atom $p$ and 
$\left\langle p|q\right\rangle = 0$ for atoms $p$ and $q$ with $p + q \leq \mathbb{I}$.

Each $a \in A_\Omega$ has a spectral decomposition
$a = s_1 e_1 + ... + s_n e_n$
with pairwise orthogonal atoms $e_k$ and $s_k \in \mathbb{R}$.
Thus $\left\langle a|a\right\rangle = s^{2}_{1} + ... + s^{2}_{n} \geq 0$
and $\left\langle a|a\right\rangle = 0$ iff $a=0$.
If $0 \leq \left\langle a|b\right\rangle$ for all $b \in A_\Omega$ with $0 \leq b$,
select $b = e_k $ and get $0 \leq s_k$ for each $k$; thus $0 \leq a$.

Now let $p_o$ be any atom. By the Riesz representation theorem
there is an element $a_o \in A_\Omega$ with 
$\mathbb{P}_{p_o}(x) = \left\langle a_o | x \right\rangle$
for all $x \in A_\Omega$.
Let $a_o = s_1 e_1 + ... + s_n e_n $ be its spectral decomposition
with pairwise orthogonal atoms $e_1, ..., e_n$ and real numbers $s_1, ..., s_n$. 
From $\mathbb{P}_{p_o}(e_j) = \left\langle a_o | e_j \right\rangle 
= \sum_{k} s_k \left\langle e_k | e_j \right\rangle = s_j$
we get $0 \leq s_j \leq 1$ for $j = 1, ..., n$.
Moreover
$1 = \mathbb{P}_{p_o}(\mathbb{I}) 
= \left\langle a_o | e_1 + ... + e_n\right\rangle
= s_1 + ... + s_n$
and $1 = \mathbb{P}_{p_o}(p_o) 
=  \left\langle a_o | p_o \right\rangle
= \sum_k s_k \left\langle e_k | p_o\right\rangle$.
The Cauchy-Schwarz inequality implicates 
$ \left|\left\langle e_k | p_o\right\rangle\right| \leq 1$ for each $k$.

For those $k$ with $s_k \neq 0$ we then get
$\left\langle e_{k} | p_o \right\rangle = 1$.
Thus $e_k$ and $p_o$ become linearly dependent. 
From $\left\|e_k\right\| =1 = \left\|p_o\right\|$ 
we get $e_k = \pm p_o$. Since $0 \leq e_{k}$ and $0 \leq p_o$
we have $e_k = p_o$.
However $e_k \neq e_j$ for $k \neq j$
and $e_k$ can coincide with $p_o$ for only one single $k_o$.
Therefore $s_k = 0$ for $k \neq k_o$, $s_{k_o} = 1$
and $a_o = e_{k_o} = p_o$.
\newpage

We now have that
$\mathbb{P}_{p}(x) = \left\langle p | x \right\rangle$
holds for each atom $p$ and
for all $x \in A_\Omega$.
For any two atoms  we get
$\mathbb{P}_{p}(q) = \left\langle p | q \right\rangle
= \left\langle q | p \right\rangle
= \mathbb{P}_{q}(p)$.
It remains to show that 
$0 \leq \left\langle a|b\right\rangle$
holds for $0 \leq a$ and  $0 \leq b$ ($a,b \in A_\Omega$).
Due to the spectral decomposition it is sufficient to prove this inequality
for atoms $p$ an $q$ and here we have already 
$\left\langle p | q \right\rangle = \mathbb{P}_{p}(q) \geq 0$.

For $x \in A_\Omega$ with the spectral decomposition 
$x = s_1 e_1 + ... + s_n e_n$ ($n \leq m$)
with pairwise orthogonal atoms $e_k$ and $s_k \in \mathbb{R}$
we get 
$\left\langle x | \mathbb{I} \right\rangle
= \sum s_k \left\langle e_k | \mathbb{I} \right\rangle
= \sum s_k \mathbb{P}_{e_k}(\mathbb{I})
= \sum s_k
= m \mu_{inv}(x)$. 
For $0 \leq x$ we have $0 \leq s_k$ for each $k$ and
$\left\|x\right\| 
= max \left\{s_1, ..., s_n \right\} 
\leq \sum s_k
=  m \mu_{inv}(x)$.
Therefore 
$\left\{ a \in A_\Omega | 0 \leq a, \mu_{inv}(a) = 1/m \right\}$ 
\linebreak $ \subseteq \left[0, \mathbb{I} \right]$.

Each atom is extreme in $\left[0, \mathbb{I} \right]$ and thus in 
$\left\{ a \in A_\Omega | 0 \leq a, \mu_{inv}(a) = 1/m \right\}$.
Vice versa, if $x$ is an extreme point in 
$\left\{ a \in A_\Omega | 0 \leq a, \mu_{inv}(a) = 1/m \right\}$,
its spectral decomposition becomes a non-trivial convex combination 
of pairwise orthogonal atoms unless $x$ itself is an atom.

For $a \in A_\Omega$ with $0 \leq a$ and $\mu_{inv}(a) = 1/m$ the map
$A_\Omega \ni x \rightarrow \left\langle a | x\right\rangle$
defines a state $\mu_a$ and 
the map $a \rightarrow \mu_a$
yields an affine isomorphism onto the state space.
The Riesz representation theorem and the self-duality 
make sure that
each state has this form.
\hfill $\square$
\vspace{0,3 cm}

With Theorem 1 we come rather close to the familiar
situation of Hilbert space quantum mechanics,
where $\mu_{inv}(x) = trace(x)/m$,
$\left\langle x|y \right\rangle = trace(xy) $
for the self-adjoint operators $x$ and $y$ (the observables) and 
where the positive operators
with normalized trace represent the (mixed) states.

The cases with information capacity $m=2$ are the generalized 
qubit models (or binary models) considered in Ref. \cite{Nie2022genqubit}.
Their state spaces are the smooth and strictly convex compact convex sets. 
If the automorphism group acts transitively on the pure states
(weak symmetry), we get the 
bit symmetry from the remarks at the end of section~3, the 
transition probability then becomes symmetric by Theorem 1
and the state space is affinely isomorphic to the unit ball in an Euclidean space,
as already shown in Ref.~\cite{Nie2022genqubit}.
We can thus conclude that the ellipsoids 
are the only smooth and strictly convex compact convex sets
where the automorphism group acts transitively on the extreme points.
In these cases, $A_\Omega$ becomes a so-called spin factor which is a special 
type of Euclidean (formally real) Jordan algebra \cite{Nie2022genqubit}.
Since the exchange symmetry holds in the spin factors, we now know that,
in the case of information capacity $m=2$
or the smooth and strictly convex compact convex sets,
the exchange symmetry is equivalent 
to the other three symmetry postulates considered here.

The above proof is partly
adopted from M\"uller and Ududec \cite{muller2012ududec},
who derive a weaker version of Theorem 1 
for a more general situation. They show that
any bit-symmetric compact convex set is self-dual.
They do not have the property ($\ast \ast$) 
with its implications like a reasonable 
atomic quantum logic and spectrality
and they do not consider the transition probability between
the atoms. Moreover their proof becomes more tricky and
must use other methods. 
A simple example of a bit-symmetric compact convex set
that does not have the property ($\ast \ast$) 
is the pentagon \cite{muller2012ududec}.

\section{Strong symmetry}

\noindent
\textbf{Corollary 1:} 
\itshape
Let $\Omega$ be any strongly symmetric finite-dimensional compact convex set
with the property ($\ast \ast$). 
Then $\Omega$ is either a simplex or the state space of a simple 
Euclidean Jordan algebra.
\normalfont
\vspace{0,3 cm}

Proof.
Since the strong symmetry implies the bit symmetry, we can apply Theorem 1 
and get the inner product $\left\langle \ | \ \right\rangle$.
Now let $\mu$ be any state. By the Riesz representation theorem
there is $a \in A_\Omega$ with
$\mu(x) = \left\langle a | x \right\rangle$ for all $x \in A_\Omega$.
The self-duality gives $0 \leq a$. Let $a = s_1 e_1 + ... + s_n e_n$
be the spectral decomposition of $a$, where $0 \leq s_k$ holds for each $k$
and the $e_k$ are pairwise orthogonal atoms (section 2 (e)). Then
$$\mu(x) = \left\langle a | x \right\rangle =\sum s_k \left\langle e_k | x \right\rangle
= \sum s_k \mathbb{P}_{e_k}(x)$$
for all $x \in A_\Omega$ and $1 = \mu(\mathbb{I}) = \sum s_k$.
From $0 \leq s_k$ we get that $\mu$ becomes the convex combination 
of the orthogonal (perfectly distinguishable) pure 
states $\mathbb{P}_{e_k}$, $k = 1, ..., n$. 
This is another type of spectrality,
which differs from (e) in section 2
and which is used by Barnum and Hilgert \cite{BarnumHilgert2020}.
We are now able to apply their theorem 
(\textit{Every strongly symmetric compact convex set with their type of 
spectrality is a simplex or the state space of a simple 
Euclidean Jordan algebra})
and get the desired result.
\hfill $\square$
\vspace{0,3 cm}

The simplexes represent the state spaces
of classical probability theory with finite dimension.
An $n$-simplex is the set 
$$\left\{ (s_1, ..., s_n) \in \mathbb{R}^{n} | 0 \leq s_1, ..., s_n, \ s_1 + ... + s_n = 1 \right\}$$
or any other set that is affinely isomorphic to this one.

In Ref. \cite{nie2023conv_self-dual} 
it was concluded at the end of section 6
that the property ($\ast \ast$), \emph{strong symmetry 
and the symmetry of the transition probability}
are possible only
with the simplexes and the state spaces of the simple 
Euclidean Jordan algebras.
Corollary 1 now reveals that the assumption
that the transition probability is symmetric
was redundant there, since this follows 
from the strong symmetry.

Thus we have almost reconstructed finite dimensional quantum theory
from the strong symmetry and the property ($\ast \ast$).
Still included is the exceptional Jordan algebra
that consists of the Hermitian $3 \times3 $-matrices 
over the octonions and that does not possess a 
representation as Hilbert space operators.
\newpage
\section{Conclusions}

The usual transition probability in Hilbert space quantum theory
is symmetric. Is there a deeper physical, probabilistic or 
information theoretical reason why nature chose this or 
are extensions of the theory with non-symmetric
transition probability conceivable? 
Here we have
revealed a connection to another kind of symmetry
and we now know that we must drop the bit symmetry
when we abandon the symmetry of the transition probability.

In quantum computation reversible
transformations from any logical bit to any other logical bit
are usually regarded as necessary and this is considered 
an information theoretic reason for the bit symmetry \cite{muller2012ududec}.
A deeper look at some quantum informational procedures,
however, shows that this rationale is 
not as clear as it seems.
Grover's search algorithm and teleportation require 
assumptions that implicate the symmetry of the transition probability,
but not the bit symmetry \cite{Nie2017TandGrover}.
The no-cloning theorem and
the quantum key distribution protocols need neither
the bit symmetry nor the general validity of the symmetry 
of the transition probability
\cite{nie2017QKD, nie2021generic}.

From the physical point of view, a continuous reversible time evolution
from one atom $p$ to any other atom $q$
(or from one pure state to any other one)
might be a crucial requirement. This means that there are
automorphisms $U_t$, $t \in \left[0,1\right]$, such that
$U_0$ is the identity, $U_1(p)=q$
and the function $t \rightarrow U_t$ is continuous.
This requirement, which has not been considered here,
implicates only the weak symmetry
among the symmetry features we have looked at.

So we must finally conclude that,
neither for the bit symmetry
(and even less for the strong symmetry)
nor for the symmetry of the transition probability,
any truly convincing physical or 
information theoretical reason is seen.
The symmetry of the transition probability
appears to be needed in some cases 
of the quantum informational procedures mentioned above,
whereas it may surprise that the bit symmetry plays 
no role there.
\small
\vspace{0,3 cm}

\noindent
\textbf{Funding}
\newline
This research received no specific grant from any funding agency in the public, commercial, or not-for-profit sectors.
\vspace{0,3 cm}

\noindent
\textbf{Conflicts of Interest}
\newline
The author declares that he has no conflicts of interest to disclose.
\vspace{0,3 cm}

\noindent
\textbf{Data Availability Statement}
\newline
No data were created or analyzed in this study.
\bibliographystyle{unsrt}
\bibliography{Literatur2024}
\end{document}